\documentclass{svjour3}                     
\smartqed  
\usepackage{graphicx}
\begin{document}

\title{Observation of exotic resonances for $K_s^0\pi^±$, $K_s^0p$ and $K_s^0\Lambda$
spectra in p+A collisions at 10 GeV/c  
}

\author{P.Zh.Aslanyan
}


\institute{Joint Institute for Nuclear Research LHEP,  Dubna, Russia
\at
              Joliot-Curie 6, Moscow region, Russia\\
              Tel.: +7-49621-65757\\
              Fax: +7-49621-65180\\
              \email{paslanian@jinr.ru}         \\
 }

\date{Received: date / Accepted: date}

\maketitle

\begin{abstract} The review on the 2m propane bubble
chamber experiment data analysis aimed to searches for an exotic
baryon states for $K^0_s$-meson subsystems. The observation of
$\Sigma^0$, $\Sigma^{*+}$(1385) and $K^{*\pm}$(892) well known
resonances  from PDG are a good tests of this method. There are
found a resonant structures for $K^0_s\pi^{\pm}$, $K^0_s$p and
$K^0_s\Lambda$ invariant mass spectra which were interpreted as
$\kappa$(720)-meson, $\Theta^+(1540)$-baryon and $N^0(1750)$ or
$\Xi^0$-baryon states, respectively.

\end{abstract}

\section{Introduction}
\label{sec:1}

 First experimental evidence for $\Theta^+$-baryon with positive strangeness
  had came  from experimental groups LEPS, Japan.
Rotational spectra of $\Theta^+\to K^0_s p$  has observed on this
experiment \cite{theta}, where significant peak in $K^0_s p$ mass
spectrum is equal to $M_{\Theta}$ = 1540$\pm$ 8 MeV/$c^2$,
$\Gamma$=(9.2$\pm$1.8) MeV/$c^2$.
 These values of $M_{\Theta}$ and $\Gamma$ are  agreed with such ones from
PDG-2004.

Recent reports for $\Theta^+$ observation are published  where
statistical significance increased for $\Theta^+ \to K^0_sp$ until
7.3 S.D. from  DIANA and 8.0 S.D. from SVD2 collaborations. An
opposite viewpoint is that all positive results might arise as
statistical fluctuations and do not reveal a true physical effect
\cite{hick}.

The scalar mesons are especially important to understand because
they have the same quantum numbers as the vacuum.  A lighter and
very broad  $\kappa$ pole is nonetheless possible and should
     be looked for in future data analyzes.
The $K^0_s\pi^{\pm}$ invariant mass
  spectra has shown   resonant structures with
  $M_{K^0_s\pi}$=720 MeV/$c^2$ and $\Gamma_e \ge 145(or
  50)$MeV/$c^2$\cite{spin06}-\cite{hs07}.

\section{$K^0_s\pi^+$ - spectrum}
\label{sec:2}

A study vector mesons $K^{*\pm}$(892) in pp interactions at 12 and
24 GeV/c by using data(280000 - events) from proton exposure of CERN
2m hydrogen bubble chamber. Total inclusive cross sections in pp
interactions are equal to 0.27$\pm$ 0.03 and 0.04$\pm
^{0.02}_{0.03}$ for $K^{*+}$ and $K^{*-}$, respectively.

\begin{figure}
{\includegraphics[width=60mm,height=35mm]{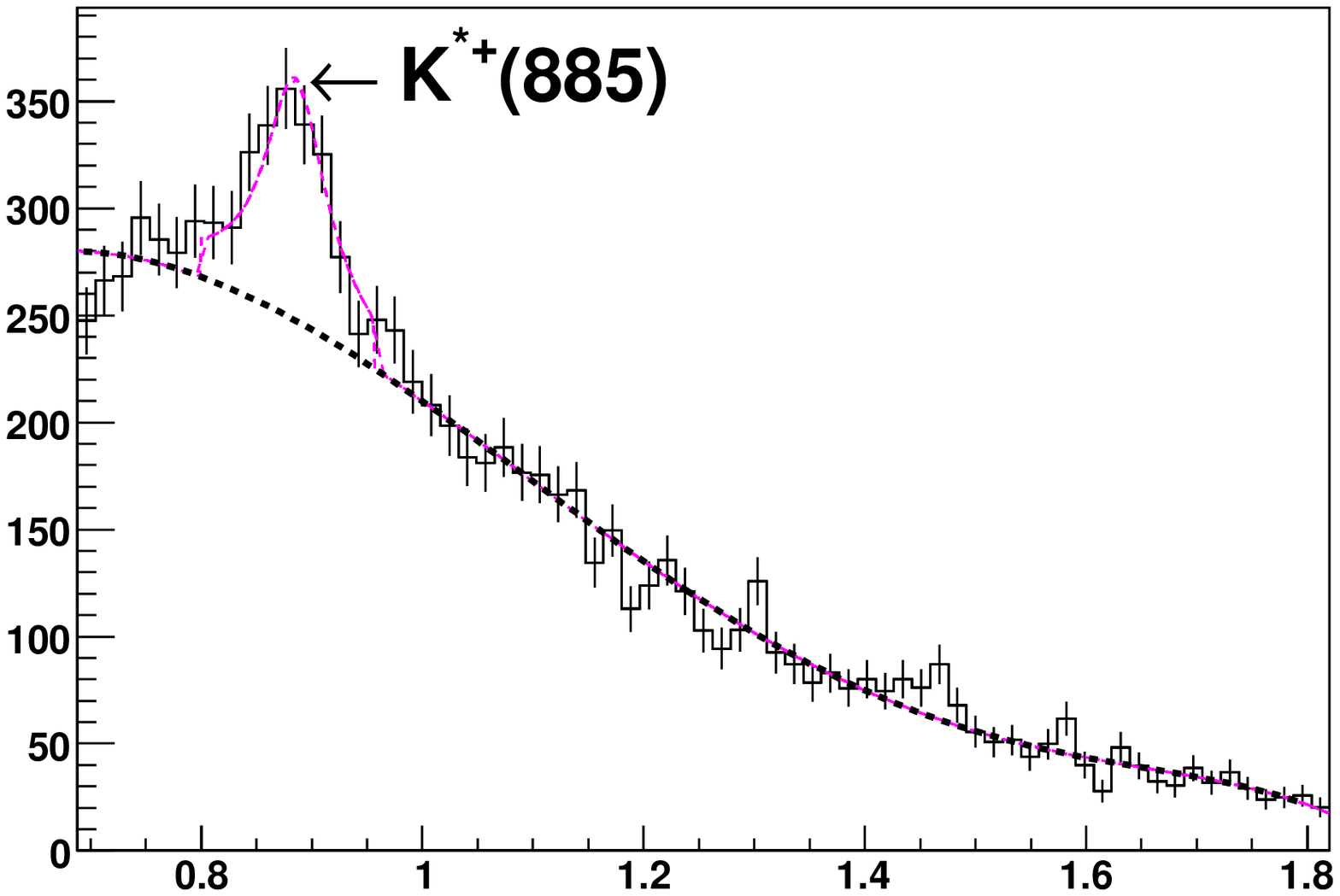}a)}
{\includegraphics[width=60mm,height=35mm]{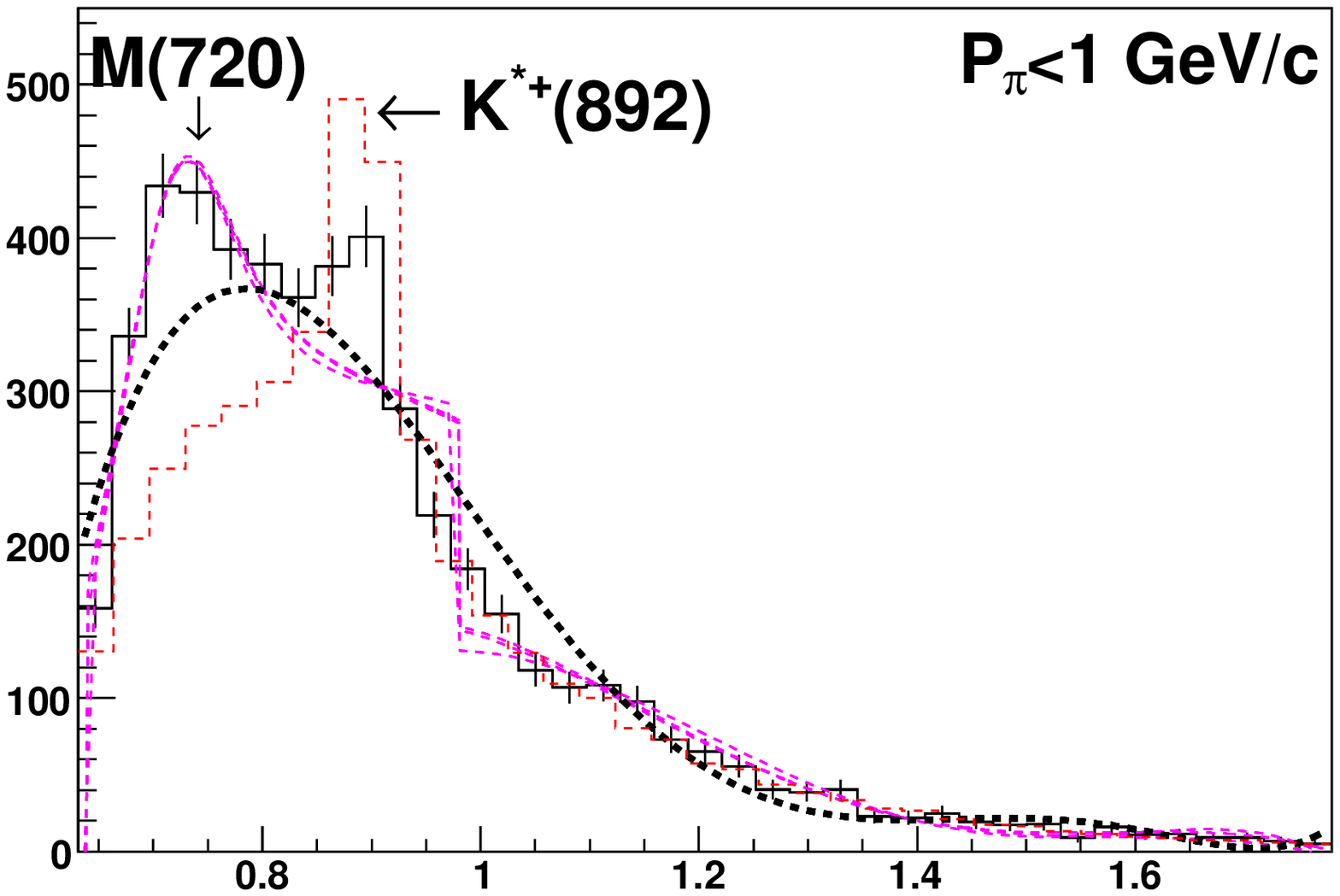}b)}
   \caption{a)All comb. for the $K^0_s \pi^+$  spectrum with bin size 16 MeV/$c^2$;
   b)The $K^0_s \pi^+$  spectrum over momentum range
  of $P_{\pi}<1 GeV/c$ with bin size 13 MeV/c$^2$. The dashed histogram is simulated events by FRITIOF.
   The dashed curve is a  background by polynomial method.
    }
  \label{kpipf}
\end{figure}

Figure~\ref{kpipf}a has  shown the effective mass distribution for
all experimental 9539($K^0_s\pi^+$ )combinations with bin sizes 16
MeV/$c^2$\cite{spin06}-\cite{hs07}. The average mass resolution for
$K^0_s\pi$ system is equal to $\approx$2\%.  The above dashed curve
in Figure~\ref{kpipf}a is the sum of a background taken in the form
of a polynomial up to the 8-th degree and 1BF
function($\chi^2/n.d.f.=73/69)$. There is significant enhancement in
mass range of 885 MeV/$c^2$, 9 S.D.,$\Gamma \approx 48$. The peak in
invariant mass spectrum at M(885) is identified as well known
$K^{*+}$(892) resonance from PDG.  The cross section of K*(892)
production (430 exp. events) is equal to 0.5 mb at 10 GeV/c for p+C
interaction. In case of bin size 13 MeV/c$^2$ there are negligible
enhancements in mass regions of: 730,780, 890 and 970
MeV/$c^{2}$\cite{spin06}-\cite{hs07}.

The effective mass of ($K^0_s\pi^+$ )distributions for 4469
combinations over the momentum range of $P_{\pi^+}<1.0$ GeV/c with
bin sizes 31 MeV/$c^2 $ are shown in Figure~\ref{kpipf}b. The
($K^0_s\pi^+$ )spectrum in Figure~\ref{kpipf}b is taken by the sum
of 8-order polynomial form and 1 BW function what is satisfactorily
described ($\chi^2/N.D.F.=43/37$)without mass range of
$K^{*+}(892)$(0.75$< M_{K^0_s\pi}<$0.98). The background by FRITIOF
or polynomial methods has approximately same form when they were
done approximation by 2BW functions\cite{spin06}. Then there are
observed significant enhancements in mass regions of: 720(7.3 S.D.)
and 890(5.5 S.D.) MeV/$c^{2}$.  After cut of $P_{\pi^+}<1.0$ GeV/c
in Figure~\ref{kpipf}b is shown that signal in mass range of 720
MeV/$c^{2}$ increased.

\subsection{$K^0_s\pi^-$ - spectrum}\label{sec:3} \indent

\begin{figure}

{\includegraphics[width=60mm,height=35mm]{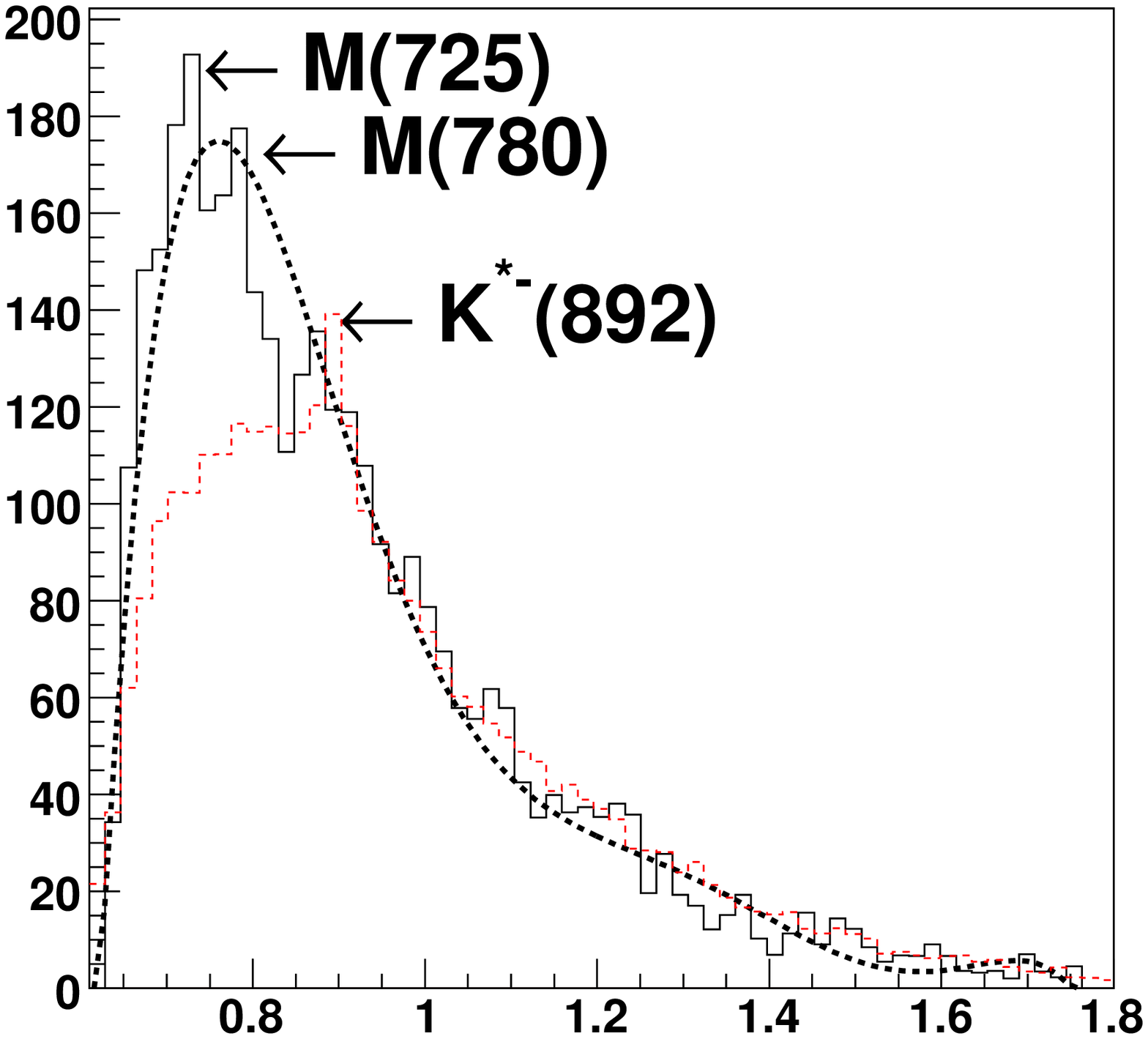}a) }
{\includegraphics[width=60mm,height=50mm]{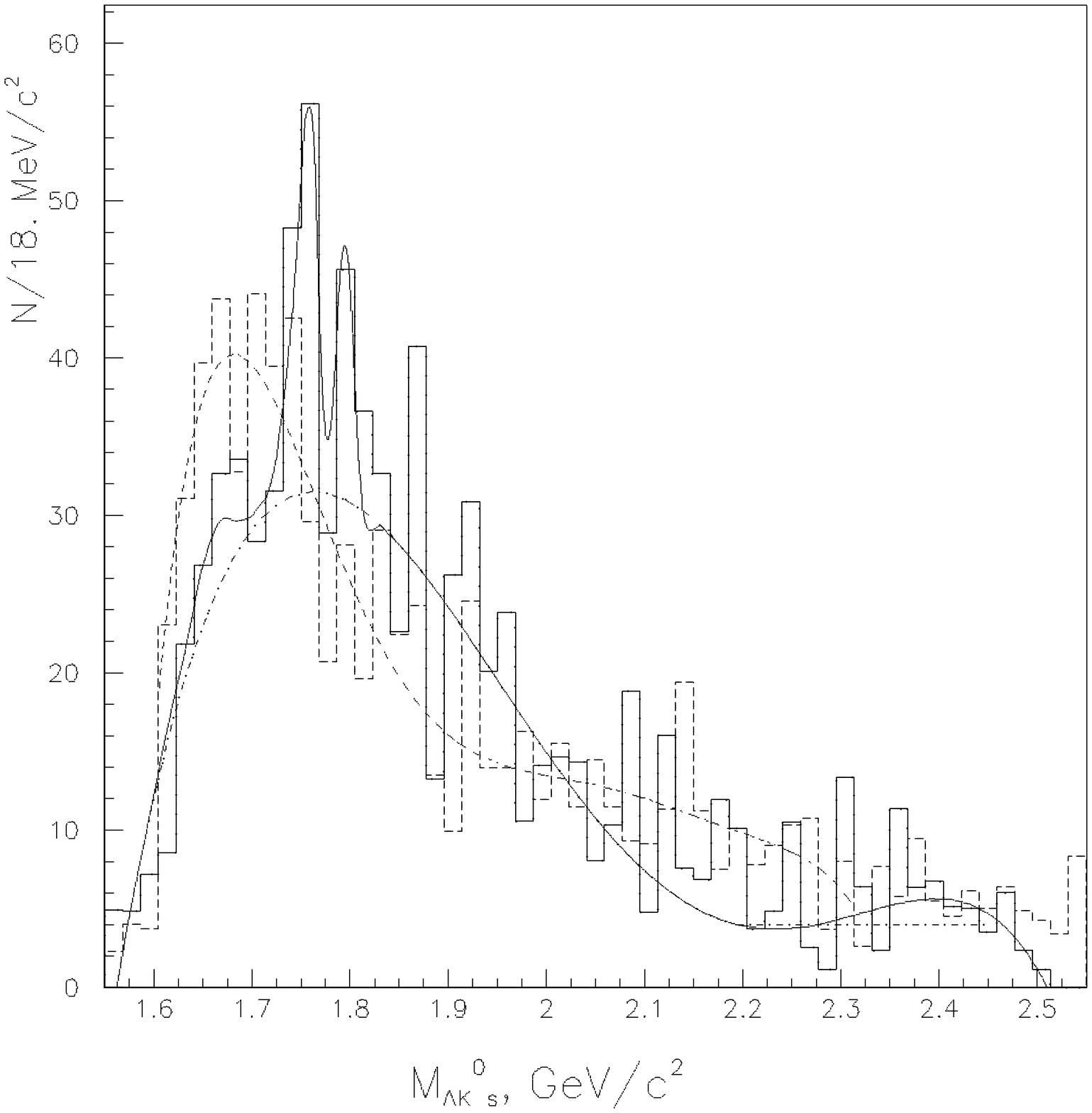}b)}
  \caption{ a)The $K^0_s \pi^-$  spectrum
  with bin size 34 MeV/c$^2$; b) The $K^0_s \Lambda$  spectrum with bin size 18
  MeV/$c^2$.
    }
  \label{kpim}
\end{figure}

Figure~\ref{kpim}a has shown the invariant mass distribution
 of 3148($K^0_s\pi^-$ )combinations with bin sizes 18 MeV/$c^2$ \cite{spin06}-\cite{hs07}.
 Figure~\ref{kpim}a has shown that the 8-order
polynomial function is approximated ($K^0_s\pi^-$ )spectrum with
$\chi^2$/n.d.f.=114/65. The sum of 1BW and background taken in the
form of a polynomial up to the 8-th degree($\chi^2=42/36$ without
mass ranges of $K^{*-}(892)$\cite{spin06}. The background by FRITIOF
or polynomial methods has approximately same form when they were
approximated with adding 2BW functions. In the ($K^0_s\pi^-$ )and
($K^0_s\pi^+$ ) spectrum there are same significant enhancements in
mass regions of 720,780,890, 980 and 1070 MeV/$c^2$(3.1
S.D.,$\approx$ 45 events)(Figure~\ref{kpim}a). The signal in mass
range of 890 MeV/c$^2$ is identified as well known resonances
$K^{*-}$(892)from PDG. The cross section of $K^{*-}(892)$ is
approximately 10 time lesser than for $K^{*+}(892)$ in this
experiment too. The preliminary total cross section for M(720) in
p+propane interactions is larger than 30$\mu$b.

\subsection{$\Lambda K^0_s$ - spectrum}\label{sec:4} \indent

Figure~\ref{kpim}b  shows the invariant mass of 1012($\Lambda
K^0_s$) combinations with bin sizes 18 MeV/$c^2 $\cite{lk}. The
solid curve is the sum of the background (detained by the first
method) and 2 Breit-Wigner  curves(Figure~\ref{kpim}b).  The
structure of mass spectrum has shown, that the significant
enhancements has been observed in two effective mass ranges 1750
MeV/$c^2$(5.6 S.D.) and 1795(3.3 S.D.) MeV/$c^2$.

These peaks could be interpreted as a possible candidates of two
pentaquark states: the $N^0$ with quark content udsds decaying into
$\Lambda K^0_s$ and the $\Xi^0$ quark content udssd decaying into
$\Lambda \overline{K^0_s}$. The preliminary total cross section for
$N^0(1750)$ production in p+propane interactions is estimated to be
$\approx 30 \mu$b.

 \section{$K^0_sp$ - spectra}
    \label{sec:5} \indent

\begin{figure}
 {\includegraphics[width=40mm,height=50mm]{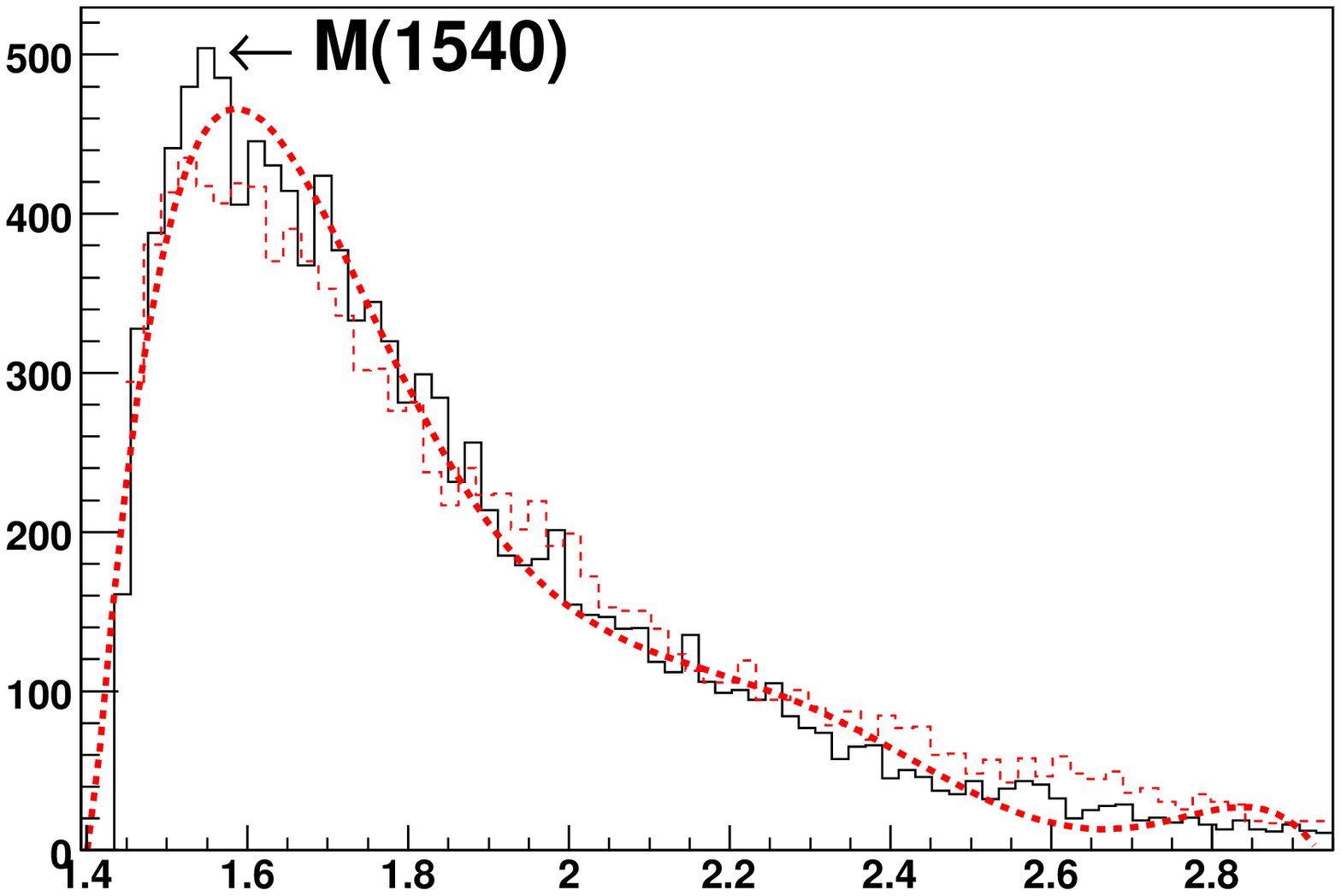}a}
 {\includegraphics[width=40mm,height=50mm]{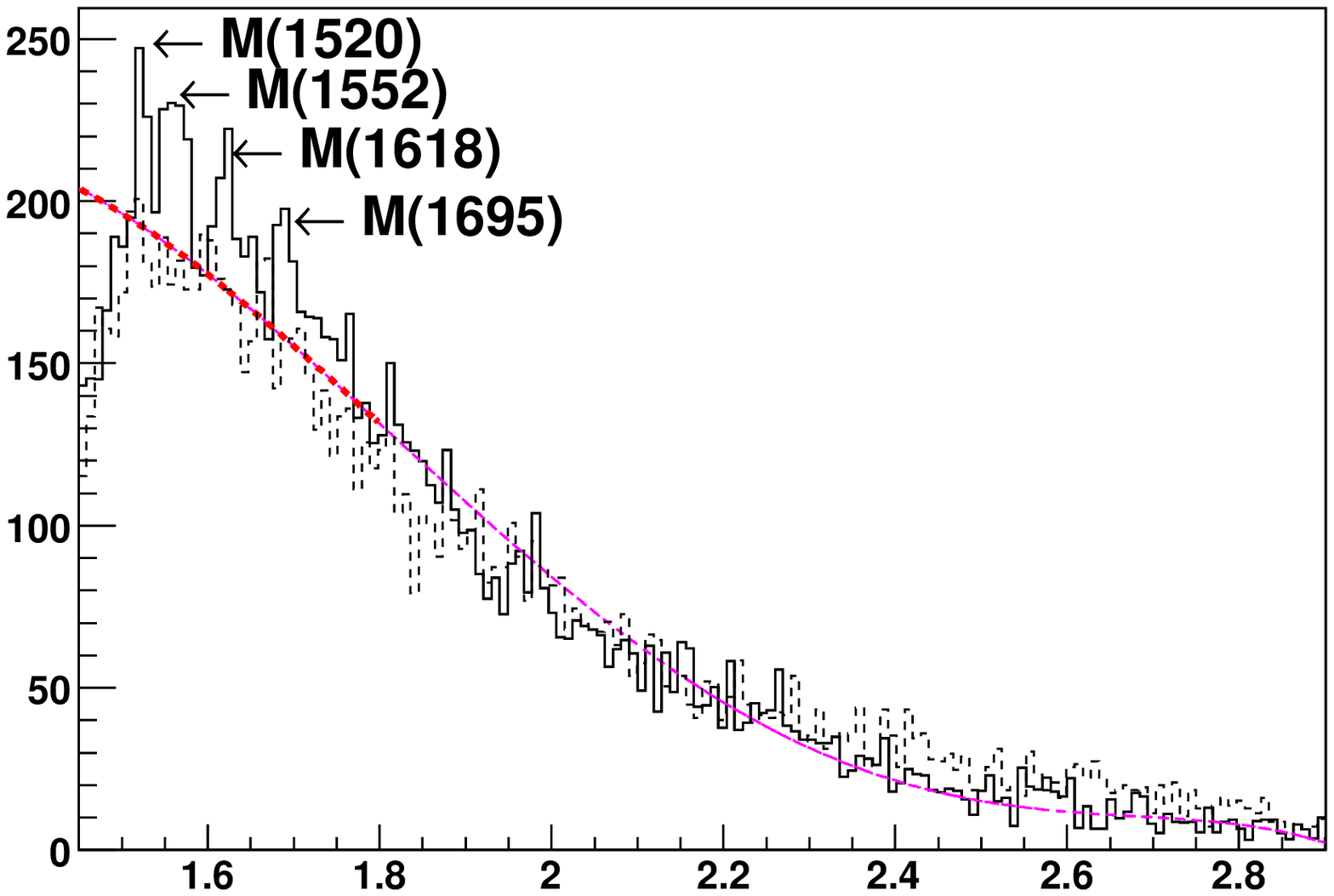}b}
{\includegraphics[width=30mm,height=65mm]{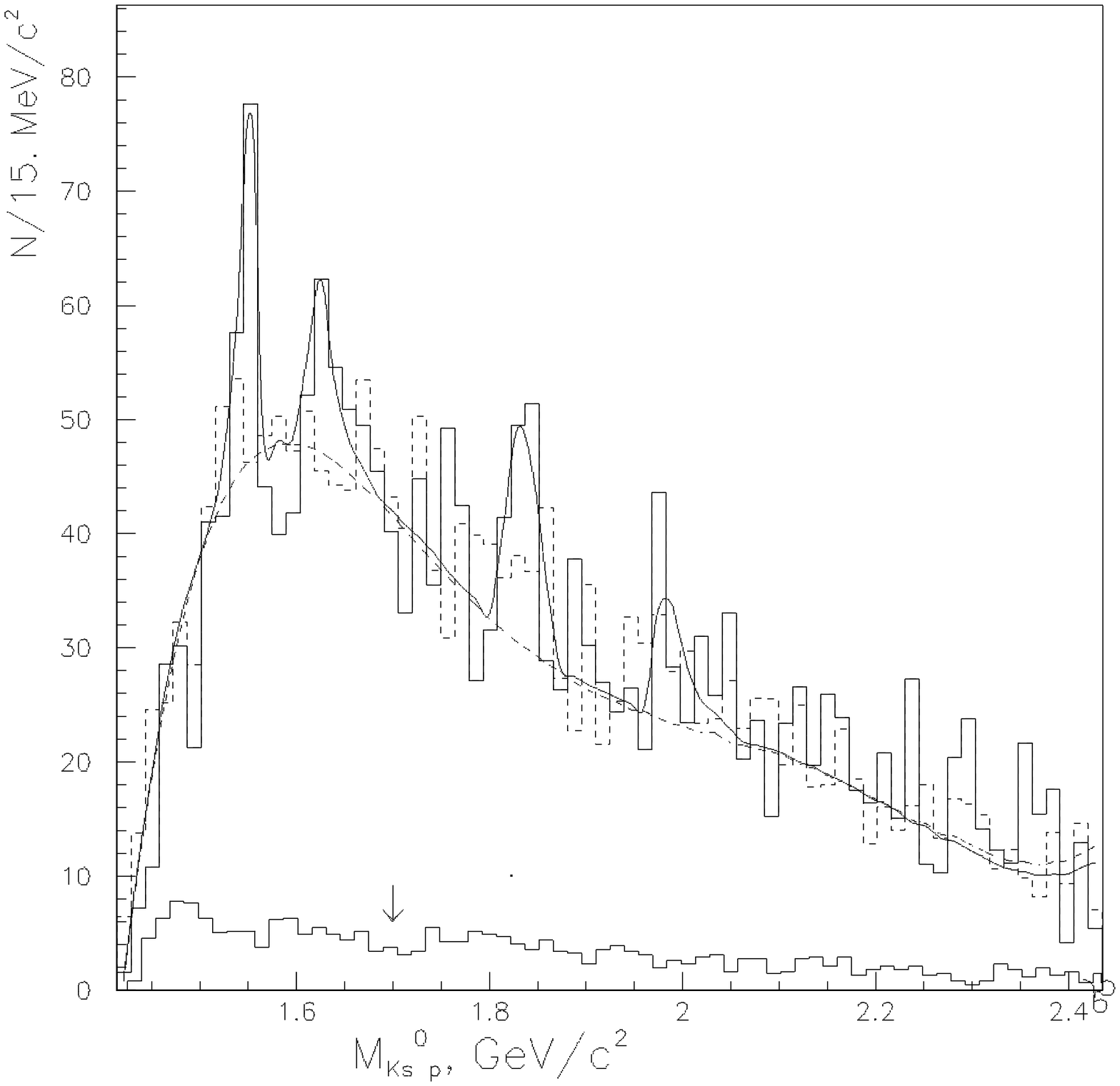}c}
  \caption{ All comb. for the $K^0_s p$  spectrum with bin sizes
  a)22 and b) 10 MeV/c$^2$;c)The $K^0_s p$  spectrum for identified protons in range of 0.35$<P_p<$0.90
  GeV/c( $\overline{K^0}p$ comb. by FRITIOF).
 }
  \label{kp}
\end{figure}

The ($K^0_s,pos. track$)  effective  mass distribution for  all
10534 combinations with bin size 22 and 10 MeV/$c^2$ are shown in
Figure~\ref{kp}a,b, respectively.  There is significant enhancement
in mass region 1540($>$5 S.D.,$\Gamma_e$=45 MeV/$c^2$) with width
$\le$ 30 MeV/$c^2$.   At bin size 10 MeV/$c^2$ the ($K_s^0p$)
effective mass spectrum has shown significant resonant structures
with M = 1520 ($\ge$4.5 S.D.,$\le$ 13 MeV/$c^2$ ),1552($\ge$ 5.9
S.D., $\le$ 15 MeV/$c^2$), 1618(3.8 S.D., $\approx$ 36 MeV/$c^2$),
and 1695 (3.8 S.D., $\approx$40 MeV/$c^2$ ). The  peak in mass range
of 1540 with width 30 MeV/$c^2$ with bin size 22 MeV/$c^2$   can
interpret as a sum of two peaks in mass ranges of 1520 and 1552
MeV/$c^2$ with widths  $<$ 15 MeV/$c^2$.  These observed peaks in
mass ranges of  1520 and 1695 can be a reflection from
$\Lambda^*$(1520) and $\Lambda^*$(1690) resonances.

 The $K^0_sp$ effective mass distribution for 2300  combinations with
 identified protons at momentum range of $0.350\le P_p\le 0.900$ GeV/c  is published
  in \cite{theta},\cite{hs07}.  The $K^0_sp$ invariant
mass spectrum shows resonant structures with $M_{K_s^0 p}$=1540(5.5
S.D.), 1613(4.8 S.D.), 1821(5.0)MeV/$c^2$.
 The experimental spectrum for $\Theta^+$ agree with the
calculated rotational spectra from the theoretical reports of D.
Akers, V.H.Mac-Gregor, A.Nambu, P.Palazzi.

\section{Conclusion} \label{sec:6} \indent

 The observation of $\Sigma^0$,$\Sigma^{*+}$(1385) and
$K^{*+}$(890)resonances are a good tests for applied method. These
interesting results for observation of $\kappa(720)$,
$\Theta^+(1540)$ and $N^0(1750) or \Xi^0$ resonances will need to
study in future experiments.

\begin{acknowledgements}
 My thanks EXA/LEAP-08 Org. Committee  for providing the excellent atmosphere during the Conference and for the financial support.
 \end{acknowledgements}

\end{document}